\documentclass[aps,prc,reprint,amsmath]{revtex4-1}

\newcommand\spn{\text{span}}
\newcommand\I{\text{\boldmath$I$}}
\newcommand\ba[4]
{| \tfrac{#1}2 \tfrac{#2}2 \tfrac{#3}2 \tfrac{#4}2 \rangle}

\begin{document}

\title{Relation to a property of the angular momentum zero space of
  states of four fermions in an angular momentum $j = 9/2$ shell
  unexpectedly found to be stationary for any rotationally invariant
  two-body interaction}

\author{K. Neerg\aa rd}

\affiliation{Fjordtoften 17, 4700 N\ae stved, Denmark}

\begin{abstract}
  The existence of states with angular momenta $I = 4$ and~6 of four
  fermions in an angular momentum $j = 9/2$ shell that are stationary
  for any rotationally invariant two-body interaction despite the
  presence of other states with the same angular momentum, the
  \textit{Escuderos-Zamick} states, is shown to be equivalent to the
  invariance to any such interaction of the span of states generated
  from $I = 0$ states by one-body operators. This invariance is
  verified by exact calculation independently of previous
  verifications of the equivalent statement. It explains the
  occurrence of the Escuderos-Zamick states for just $I = 4$ and~6.
  The action of an arbitrary interaction on the invariant space and
  its orthogonal complement is analyzed, leading to a relation of the
  Escuderos-Zamick energy levels to levels with $I = 10$ and~12. Parts
  of the observed spectra of $^{72}$Ni, $^{74}$Ni, $^{94}$Ru,
  $^{96}$Pd, $^{212}$Rn, and $^{214}$Ra are discussed in the light of
  this relation.
\end{abstract}

\maketitle

\section{\label{sec:intr}Introduction}

Escuderos and Zamick found in a numeric study of the system of four
nucleons in an angular momentum $j = 9/2$ shell of a semi-magic
nucleus that for each angular momentum $I = 4$ and~6, the states in
one irreducible module of the angular momentum algebra, briefly a
\textit{multiplet}, are stationary for any rotationally invariant
two-body interaction~\cite{ref:Esc06}, that is, \textit{solvable} in
Talmi's terminology~\cite{ref:Tal10}, despite the presence of other
multiplets with the same $I$. It follows that they have definite
seniority $v$~\cite{ref:Tal10}, which gives rise, in certain nuclei,
to particular patterns of transition rates in E2 decay and
single-nucleon transfer~\cite{ref:Isa08,ref:Isa14,ref:Qi17}. The
solvability of the Escuderos-Zamick states was subsequently confirmed
in exact calculations by Van Isacker and
Heinze~\cite{ref:Isa08,ref:Isa14}, and Qi, Xu and
Liotta~\cite{ref:Qi12}. These calculation are case-by-case
examinations of the individual instances of two-body and total angular
momentum, which led the authors of \cite{ref:Isa14} to conclude that
``a simple, intuitive reason for [the solvability] is still lacking''.
I show below that the existence of the Escuderos-Zamick states is
equivalent to a property of the space of $I = 0$ states of the system.
The verification of this property again leads to an examination of
several cases one by one. The equivalence explains, however, that the
solvable multiplets occur for exactly $I = 4$ and~6.

Throughout this paper, $j = 9/2$. Let $\Phi_0$ denote the space of
$I = 0$ states of the four-fermion system, and let $a_m$ be the
annihilator of a fermion in the state $|jm\rangle$ in the conventional
notation~\cite{ref:Edm57}. One can then define a space
\begin{equation}\label{eq:Phi4}
  \Phi_4 = \spn_{m,m'} \, a_m^\dagger a_{m'} \Phi_0 .
\end{equation}
The property to be verified below and shown there to be equivalent to
the existence of the Escuderos-Zamick states is the following.
\textit{$\Phi_4$ is invariant to any rotationally invariant two-body
  interaction}. To see how this explains the appearance of solvable
multiplets for just $I = 4$ and~6, note that the tensor operators
$T_{IM_I} = \sum_{mm'} (-)^{j-m'} \linebreak \langle j m j -m' | I M_I
\rangle a_m^\dagger a_{m'}$, where
$\langle j_1 m_1 j_2 m_2 | j m \rangle$ is the vector coupling
coefficient~\cite{ref:Edm57}, form a basis for the span of operators
$a_m^\dagger a_{m'}$. The subspace of $\Phi_4$ carrying quantum
numbers $I,M_I$ is $T_{IM_I} \Phi_0$. Now consider
Table~\ref{tbl:mult}, obtained by a straightforward count of
$m$-combinations.
\begin{table}
  \caption{\label{tbl:mult}Multiplicities of multiplets of four
    angular momentum $j = 9/2$ fermions per angular
    momentum $I$ and seniority $v$.}
\begin{ruledtabular}
\begin{tabular}{c|cccccccccccc}
$v$&\hskip-15pt$I$\hskip1pt\vrule&0&2&3&4&5&6&7&8&9&10&12\\\hline
0&&1&0&0&0&0&0&0&0&0&0&0\\
2&&0&1&0&1&0&1&0&1&0&0&0\\
4&&1&1&1&2&1&2&1&1&1&1&1
\end{tabular}\end{ruledtabular}
\end{table}
Since $\Phi_0$ is 2-dimensional, $T_{IM_I} \Phi_0$ has dimension 2, at
most. Angular momenta $I = 4$ and~6 are the only ones allowing more
than 2 linearly independent multiplets in the four-fermion system,
exactly 3 in both cases. It may be verified by direct calculation, and
also follows from a general result in Sec~\ref{seq:Phi}, that in each
case, $T_{IM_I} \Phi_0$ is exactly 2-dimensional. If $\Phi_4$ is
invariant to a Hermitian and rotationally invariant operator $V$, then
so is also $T_{IM_I} \Phi_0$, and so is then also its 1-dimensional
orthogonal complement within the space of states with quantum numbers
$I,M_I$. This means that the states in the orthogonal complement are
eigenstates of $V$.

The proof of equivalence is completed in Sec.~\ref{sec:anal}, and the
verification of the invariance of $\Phi_4$ in Sec.~\ref{seq:V}.
Analyzing the actions of an arbitrary $V$ on $\Phi_4$ and its
orthogonal complement $\Phi_4^\perp$ reveals remarkable regularities,
one of which leads to a rule for relative level spacings that is
accessible to experimental verification and so far lacks fundamental
explanation. This analysis is the topic of Secs.~\ref{sec:actPhiperp}
and \ref{sec:actPhi}, followed by my conclusion in Sec.~\ref{sec:con}.
A detail of my formalism is discussed and one other observed
regularity explained in two appendices.

\section{\label{sec:anal}Analysis}

Below, $I_\text{E} = 4$ or~6. Important for the following is also the
space $\Phi_3$ of states with $I = j$ of three $j = 9/2$ fermions,
which is spanned by single multiplets $\Phi_{3v}$ with $v = 1$ and 3.
For each $v$, at most one multiplet with a given $I$ can be formed by
adding a $j = 9/2$ fermion to the states in $\Phi_{3v}$. This
multiplet can be written $P_I \, \spn_m \, a_m^\dagger \Phi_{3v}$,
where $P_I$ is the projection onto angular momentum $I$. It may be
verified by direct calculation, and also follows from the general
result in Sec~\ref{seq:Phi}, that for each $I_\text{E}$, these two
multiplets are independent. The space
$\Phi_{I_\text{E}\gamma} = P_{I_\text{E}} \, \spn_m \, a_m^\dagger
\Phi_{31}$ necessarily has $v = 2$. By Table~\ref{tbl:mult}, its
orthogonal complement $\Phi_{I_\text{E}\beta}$ within
$P_{I_\text{E}} \, \spn_m \, a_m^\dagger \Phi_3$ then has $v = 4$. The
Escuderos-Zamick multiplet $\Phi_{I_\text{E}\alpha}$ is, by
definition, the orthogonal complement of $\Phi_{I_\text{E}\beta}$
within the space of states with $I = I_\text{E}$ and $v = 4$ of the
four fermions. It may be characterized also among such states by
vanishing parentage by $\Phi_{33}$~\cite{ref:Esc06}. Evidently, it is
also the orthogonal complement of
$P_{I_\text{E}} \, \spn_m \, a_m^\dagger \Phi_3$ within the space of
states with $I = I_\text{E}$.

In the remainder of this paper, $V$ denotes any rotationally invariant
two-body interaction. Because $V$ acts as a scalar on the irreducible
module $\Phi_{I_\text{E}\alpha}$, the states in
$\Phi_{I_\text{E}\alpha}$ being eigenstates of $V$ is equivalent to
$\Phi_{I_\text{E}\alpha}$ being invariant to $V$. By Hermiticity of
$V$ and conservation of angular momentum, this is, in turn, equivalent
to $P_{I_\text{E}} \, \spn_m \, a_m^\dagger \Phi_3$ being invariant to
$V$. The space $\spn_m \, a_m^\dagger \Phi_3$ cannot contain states
with $I > 2j$. For every $I \le 2j$ except $I = 4$ and~6, it may be
verified by direct calculation, and also follows from the general
result in Sec~\ref{seq:Phi}, that
$P_I \, \spn_m \, a_m^\dagger \Phi_3$ exhausts the space of states of
the four fermions with angular momentum $I$ and thus is invariant to
any rotationally invariant operator. Invariance of both spaces
$P_{I_\text{E}} \, \spn_m \, a_m^\dagger \Phi_3$ to $V$ is then
equivalent to $\Phi_4 = \spn_m \, a_m^\dagger \Phi_3$ being invariant
to $V$. In summary, the existence of the Escuderos-Zamick states is
equivalent to $\Phi_4$ being invariant to any $V$.

To establish the equivalence stated in the introduction, it remains to
show that $\Phi_4$ can be written in the form~\eqref{eq:Phi4}. To this
end, notice $\spn_m \, a_{m} \Phi_0 \subset \Phi_3$. It may be
verified by direct calculation, and also follows from the general
result in Sec~\ref{seq:Phi}, that the left hand side exhausts $\Phi_3$
so that $\spn_m \, a_{m} \Phi_0 = \Phi_3$. This evidently leads to the
expression~\eqref{eq:Phi4}. The remainder of this paper is dedicated
to a proof (independent of the proofs
in~\mbox{\cite{ref:Isa08,ref:Isa14,ref:Qi12}} of the equivalent
statement) that $\Phi_4$ as given by~\eqref{eq:Phi4} is actually
invariant to any $V$, and analyses of the actions of an arbitrary $V$
on $\Phi_4$ and its orthogonal complement.

\section{\label{seq:Phi}Spaces $\Phi_0$ and $\Phi_4$}

The structure of multi-fermion states in the $j = 9/2$ shell is
conveniently described in terms of creation operators
\begin{equation}\label{eq:alpha}
  \alpha_m^\dagger = \sqrt{\frac{(j+m)!}{(j-m)!}} a_m^\dagger ,
\end{equation}
corresponding to unnormalized single-fermion states. In terms of the
usual complex coordinates $(I_0,I_\pm)$ of the total angular
momentum $\I$~\cite{ref:Edm57}, these operators obey
\begin{equation}\label{eq:com}\begin{gathered}[]
  [ I_0 , \alpha_m^\dagger ] = m \alpha_m^\dagger , \\
  [ I_+ , \alpha_m^\dagger ] = \begin{cases}
      \alpha_{m+1}^\dagger , & m < j , \\
      0 , & m = j ,
    \end{cases} \\
  [ I_- , \alpha_m^\dagger ] = \begin{cases}
      (j+m)(j-m+1) \alpha_{m-1}^\dagger , & m > -j , \\
      0 , & m = -j .
    \end{cases}
\end{gathered}\end{equation}
A state of four $j = 9/2$ fermions can be expanded on the states
\begin{equation}\label{eq:ba}
  | m_1 m_2 m_3 m_4 \rangle
  = \left( \prod_{i=1}^4 \alpha_{m_i}^\dagger \right) |\rangle
\end{equation}
with $j \ge m_1 > m_2 >m_3 > m_4 \ge -j$, where $|\rangle$ is the
vacuum. The eigenspaces of $I_0$ with eigenvalues $M_I$ are spanned by
the states with $\sum_i m_i = M_I$. The space $\Phi_0$ is the subspace
of the $M_I = 0$ space characterized by $I_+ \Phi_0 = 0$. Since there
are 18 states $| m_1 m_2 m_3 m_4 \rangle$ with $M_I = 0$ and 16 with
$M_I = 1$ (in accordance with the total multiplicities for $I \ge 0$
and 1 in Table~\ref{tbl:mult}), this constraint can be expressed by a
homogeneous system of 16 linear equations in 18 expansion
coefficients. The equations turn out independent in accordance with
the dimension 2 of $\Phi_0$. Two linearly independent solution are
\begin{equation}\label{eq:phi}\begin{aligned}
  \phi_0 &= \ba{9}{7}{-7}{-9}
    - \ba{9}{5}{-5}{-9}
    + \ba{9}{3}{-3}{-9}
    - \ba{9}{1}{-1}{-9} \\
    &+ \ba{7}{5}{-5}{-7}
    - \ba{7}{3}{-3}{-7}
    + \ba{7}{1}{-1}{-7}
    + \ba{5}{3}{-3}{-5} \\
    &- \ba{5}{1}{-1}{-5}
    + \ba{3}{1}{-1}{-3} , \\
  \phi_1 &= -5 \ba{9}{7}{-7}{-9}
    +5 \ba{9}{5}{-5}{-9}
    + \ba{9}{3}{-3}{-9} \\
    &-7 \ba{9}{1}{-1}{-9}
    +9 \ba{7}{5}{-5}{-7}
    -3 \ba{7}{3}{-3}{-7} \\
    &-9 \ba{7}{1}{-1}{-7}
    -6 \ba{5}{3}{-3}{-5} \\
    &-14 \ba{9}{3}{-5}{-7}
    -6 \ba{7}{5}{-3}{-9}
    +16 \ba{9}{1}{-3}{-7} \\
    &+6 \ba{7}{3}{-1}{-9}
    -25\ba{9}{-1}{-3}{-5}
    -6 \ba{5}{3}{1}{-9} \\
    &+9 \ba{7}{1}{-3}{-5}
    +6 \ba{5}{3}{-1}{-7} .
\end{aligned}\end{equation}
Here, $\phi_0$ evidently has $v = 0$. The state $\phi_0 + 2 \phi_1$ is
orthogonal to $\phi_0$ and thus has $v = 4$. In the expansion of
$\phi_1$, the coefficients of $\ba{7}{5}{-3}{-9}$ and
$\ba{7}{3}{-1}{-9}$ are equal except for opposite signs. This is
because in the expansion of $I_+ \phi_1$, the coefficient of
$\ba{7}{5}{-1}{-9}$ gets contributions only from these two
coefficients. Similar comparisons explain that all the four states
$\ba{7}{5}{-3}{-9}$, $\ba{7}{3}{-1}{-9}$, $\ba{5}{3}{1}{-9}$, and
$\ba{5}{3}{-1}{-7}$ have equal coefficients except for a sign. The
ratios of coefficients of $|m_1 m_2 m_3 m_4\rangle$ and
$|-m_4,-m_3,-m_2,-m_1\rangle$ are \linebreak
$\prod_{i=1}^4 (j-m_i)!/(j+m_i)!$ so that the corresponding ratios in
the basis of states
$\left( \prod_{i=1}^4 a_{m_i}^\dagger \right) |\rangle$ equal one, as
required by the symmetry under half-turn rotations about axes
perpendicular to the quantization axis.

Since $\Phi_4$ is rotationally invariant, its invariance to $V$ is
equivalent to invariance of its $M_I = 0$ subspace
\begin{equation}\label{eq:Phi40}
  \Phi_{40} = \spn_m \, n_m \Phi_0 ,
\end{equation}
where $n_m = a_m^\dagger a_m$. This space is spanned by the 20 states
$n_m \phi_i$ with $m = j,j-1,\dots,-j$ and $i \in \{0,1\}$. Each of
these states is obtained by selecting in the expansion~\eqref{eq:phi}
of $\phi_i$ the terms where the orbit $|jm\rangle$ is occupied. Not
all of them are linearly independent. Thus evidently
$n_m \phi_0 =n_{-m} \phi_0$. Further, $I_0 \phi_1 = 0$ is a linear
relation among the 10 states $n_m \phi_1$. There remain 14 states,
which turn out linearly independent. This number coincides with the
total multiplicity for $I \le 2j$, excepting the Escuderos-Zamick
multiplets. The $M_I = 0$ state in every remaining multiplet thus
belongs to $\Phi_{40}$. Consequently, every such multiplet is
contained in $\Phi_4$. This requires, in turn, that the multiplets
$P_{I_\text{E}} \, \spn_m \, a_m^\dagger \Phi_{3v}$ with $v = 1$ and 3
be independent, that for $I \le 2j$ and $I \ne 4,6$, the space
$P_I \, \spn_m \, a_m^\dagger \Phi_3$ exhaust the space of states with
angular momentum $I$, and that equality hold in the inclusion
$\spn_m \, a_{m} \Phi_0 \subset \Phi_3$, all of which was used in
Secs.~\ref{sec:intr} and \ref{sec:anal}.

I choose in $\Phi_{40}$ a basis $(\psi_i | i = 1,\dots,14)$, where
$(\psi_i | i = 1,\dots,5)$ are the states $n_m \phi_0$ with
$m = j, j-1,\dots,1/2$ in this order, and $(\psi_i | i = 6,\dots,14)$
are the states $n_m \phi_1$ with $m = j,j-1,\dots,-j+1$ in this order.
By Hermiticity and angular momentum conservation, $\Phi_{40}$ is
invariant to $V$ if and only if its orthogonal complement
$\Phi_{40}^\perp$ within the $M_I = 0$ space is so. This space is
spanned by the states

\newpage

\noindent\begin{minipage}{\columnwidth}

\newcommand\qqquad{\phantom{\text{$+\frac1{000} \Big(\ $}}}
\newcommand\qqqquad{\phantom{\text{$+\frac1{0000} \Big(\ $}}}
\begin{equation}\label{eq:chi}\begin{aligned}
  \chi_1 =\ & 14\ba{9}{3}{-5}{-7} + 6\ba{7}{5}{-3}{-9} \\
    &+\frac{2}{153} \Big(
    -768\ba{9}{7}{-7}{-9}
    +231\ba{9}{5}{-5}{-9} \\
    &\qqquad+927\ba{9}{3}{-3}{-9}
    -72\ba{9}{1}{-1}{-9} \\
    &\qqquad+159\ba{7}{5}{-5}{-7}
    -553\ba{7}{3}{-3}{-7} \\
    &\qqquad+56\ba{7}{1}{-1}{-7}
    -640\ba{5}{3}{-3}{-5} \\
    &\qqquad-712\ba{5}{1}{-1}{-5}
    -840\ba{3}{1}{-1}{-3} \Big) , \\
  \chi_2 =\ & 16\ba{9}{1}{-3}{-7} + 6\ba{7}{3}{-1}{-9} \\
    &+\frac{16}{1071} \Big(
    516\ba{9}{7}{-7}{-9}
    -105\ba{9}{5}{-5}{-9} \\
    &\qqqquad-171\ba{9}{3}{-3}{-9}
    +450\ba{9}{1}{-1}{-9} \\
    &\qqqquad+345\ba{7}{5}{-5}{-7}
    +1225\ba{7}{3}{-3}{-7} \\
    &\qqqquad+364\ba{7}{1}{-1}{-7}
    +430\ba{5}{3}{-3}{-5} \\
    &\qqqquad+880\ba{5}{1}{-1}{-5}
    +966\ba{3}{1}{-1}{-3} \Big) , \\
  \chi_3 =\ & 25\ba{9}{-1}{-3}{-5} + 6\ba{5}{3}{1}{-9} \\
    &+\frac{25}{1071} \Big(
    834\ba{9}{7}{-7}{-9}
    +2184\ba{9}{5}{-5}{-9} \\
    &\qqqquad+1629\ba{9}{3}{-3}{-9}
    +279\ba{9}{1}{-1}{-9} \\
    &\qqqquad-750\ba{7}{5}{-5}{-7}
    -847\ba{7}{3}{-3}{-7} \\
    &\qqqquad-931\ba{7}{1}{-1}{-7}\
    -376\ba{5}{3}{-3}{-5} \\
    &\qqqquad-3310\ba{5}{1}{-1}{-5}\
    -2100\ba{3}{1}{-1}{-3} \Big) , \\
  \chi_4 =\ & 9\ba{7}{1}{-3}{-5} + 6\ba{5}{3}{-1}{-7} \\
    &+\frac{3}{119} \Big(
    -150\ba{9}{7}{-7}{-9}
    -231\ba{9}{5}{-5}{-9} \\
    &\qqquad-162\ba{9}{3}{-3}{-9}
    -81\ba{9}{1}{-1}{-9} \\
    &\qqquad+45\ba{7}{5}{-5}{-7}
    +196\ba{7}{3}{-3}{-7} \\
    &\qqquad+301\ba{7}{1}{-1}{-7}\
    +232\ba{5}{3}{-3}{-5} \\
    &\qqquad+508\ba{5}{1}{-1}{-5}
    +126\ba{3}{1}{-1}{-3} \Big) . 
\end{aligned}\end{equation}
It is straightforward to check $\psi_i^\dagger \chi_k = 0$ for every
$i,k$ using
$\||m_1m_2m_3m_4\rangle\|^2 = \prod_{i=1}^4 (j+m_i)!/(j-m_i)!$. For
example,
\begin{equation}\begin{aligned}
  \psi_6^\dagger \chi_1 &= \phi_1^\dagger n_{9/2} \chi_1
    = (-14)\!\cdot\!\frac{3}{7}\!\cdot\!14 \\
      &+\frac2{153} \Big(
        (-5)\!\cdot\!(-786) + 5\!\cdot\!231\\&\qqquad
        +1\!\cdot\!927 + (-7)\!\cdot\!(-72)\Big) = 0 .
\end{aligned}\end{equation}

\end{minipage}

\section{\label{seq:V}Interaction $V$ and invariance of
  $\Phi_4$}

Every $V$ is a linear combination of five basic interactions $V_J$,
where $J = 0,2,\dots,2j-1$. They can be chosen in the form
\begin{equation}\label{eq:V}
  V_J = \tfrac12 \sum_{M=-J}^J P_{JM}^\dagger P_{JM}
\end{equation}
with
\begin{equation}\label{eq:P+a}
  P_{JM} = c_J \sum_{m_1+m_2=M}
    \langle j m_1 j m_2 | J M \rangle a_{m_2} a_{m_1} ,
\end{equation}
where $c_J$ is a positive constant. I set
\begin{multline}\label{eq:cm1m2}
  c_J \langle j m_1 j m_2 | J M \rangle \\
    = \binom{2J}{J+M}^{-1/2}
      \sqrt{\frac{(j+m_1)!(j+m_2)!}{(j-m_1)!(j-m_2)!}}
      c^J_{m_1m_2} ,
\end{multline}
so that by~\eqref{eq:alpha},
\begin{equation}\label{eq:P+al}
  P_{JM}^\dagger = \binom{2J}{J+M}^{-1/2} \sum_{m_1+m_2=M}
    c^J_{m_1m_2} \alpha_{m_1}^\dagger  \alpha_{m_2}^\dagger .
\end{equation}
The definition~\eqref{eq:cm1m2} implies $c^J_{m_1m_2} = 0$ for
$|m_1 + m_2| > \linebreak J$. It follows from $[I_+,P_{JJ}^\dagger] = 0$,
\eqref{eq:P+al}, \eqref{eq:com}, and \eqref{eq:cm1m2} that $c_J$ can
be chosen such that $c^J_{m_1m_2} = (-1)^{j-m_1}$ for $m_1 + m_2 = J$.
From $[I_-,P_{JM}^\dagger] = \sqrt{(J+M)(J-M+1)} P_{J,M-1}^\dagger$
for $M > -J$, \eqref{eq:P+al}, and \eqref{eq:com}, one gets the
recursion relation
\begin{multline}\label{eq:rec-}
  (J - m_1 - m_2) c^J_{m_1m_2}
    = (j - m_1)(j + m_1 + 1) c^J_{m_1+1,m_2} \\
      + (j - m_2)(j + m_2 + 1) c^J_{m_1,m_2+1} ,
\end{multline}
which then determines $c^J_{m_1m_2}$ for $-J \le m_1 + m_2 < J$.
(Continuation of the recursion in fact results in $c^J_{m_1m_2} = 0$
for $m_1 + m_2 < -J$. Terms in~\eqref{eq:rec-} with $m_1$ or $m_2$
equal to $j$, which involve undefined values of $c^J_{m_1'm_2'}$, are
just omitted.) All $c^J_{m_1m_2}$ turn out integral, which is
explained in Appendix~\ref{sec:cmm}. From the definition~\eqref{eq:cm1m2}
and symmetries of the vector coupling coefficients~\cite{ref:Edm57},
one gets
\begin{equation}\label{eq:csym}\begin{gathered}
  c^J_{m_1m_2} = -c^J_{m_2m_1} , \\
  c^J_{m_1m_2}
  = \frac{(j-m_1)!(j-m_2)!}{(j+m_1)!(j+m_2)!} c^J_{-m_2,-m_1} ,
\end{gathered}\end{equation}
whence by \eqref{eq:P+al} and \eqref{eq:alpha} follows
\begin{multline}\label{eq:P}
   P_{JM} = \binom{2J}{J+M}^{-1/2} \\ \sum_{m_1+m_2=M}
       c^J_{m_1m_2} \frac{(j+m_1)!(j+m_2)!}{(j-m_1)!(j-m_2)!}
       \alpha_{m_2}  \alpha_{m_1} \\
     = \binom{2J}{J+M}^{-1/2} \sum_{m_1+m_2=M}
       c^J_{-m_2,-m_1} \alpha_{m_2}  \alpha_{m_1}
\end{multline}
in terms of annihilation operators
\begin{equation}
  \alpha_m = \sqrt{\frac{(j-m)!}{(j+m)!}} a_m
\end{equation}
obeying
\begin{equation}
  \{ \alpha_m , \alpha_{m'}^\dagger \} = \delta_{m,m'} .
\end{equation}

It follows that the action of $V_J$ on a basic state
$|m_1 m_2 m_3 m_4 \rangle$ can be described by the following operation
$u^{pq}_{Jm}$. If $m_p + m_q - m$ is outside the range of $m$'s, then
$u^{pq}_{Jm} |m_1 m_2 m_3 m_4 \rangle = 0$. Otherwise replace $m_p$
and $m_q$ by $m$ and $m_p + m_q - m$. If this results in two $m$'s
being equal, $u^{pq}_{Jm} |m_1 m_2 m_3 m_4 \rangle = 0$. Otherwise
reorder, if necessary, the $m$'s to decreasing order and multiply the
state by the sign of the permutation. Finally multiply the state by
$\binom{2J}{J+m_p+m_q}^{-1} c^J_{m,m_p+m_q-m} c^J_{-m_q,-m_p}$. Then
\begin{equation}
  V_J |m_1 m_2 m_3 m_4 \rangle
  = \sum_{1 \le p < q \le 4,m} u^{pq}_{Jm} |m_1 m_2 m_3 m_4 \rangle .
\end{equation}
The state $V_J \chi_i$ is obtained by applying this formula to each
term in the expansion~\eqref{eq:chi} of $\chi_i$. I did this
calculation for every $J,i$ and found that in every case, $V_J \chi_i$
is a linear combination of $\{ \chi_i | i = 1,\dots,4 \}$. This proves
that $\Phi_{40}^\perp$, and in turn, $\Phi_{40}$, $\Phi_4$, and the
orthogonal complement $\Phi_4^\perp$ of the latter, are invariant to
every $V$. For completeness, I also verified directly that every
$V_J \psi_i$ is a linear combination of
$\{ \psi_i | i = 1,\dots,14 \}$.

\section{\label{sec:actPhiperp}Action of  $V$ on $\Phi_{40}^\perp$}

The expansion of $V_J \chi_i$ on states $\chi_k$ may be expressed by a
matrix $V^{\perp J} = ( v^{\perp J}_{ik} | i,k = 1,\dots,4)$ defined
by
\begin{equation}\label{eq:VJperp}
  V_J \chi_i = \sum_{k=1}^{4} v^{\perp J}_{ki} \chi_k .
\end{equation}
These matrices are given by
\begin{equation}\label{eq:VJperpmat}\begin{aligned}
  V^{\perp0} &= \left(\begin{smallmatrix}
    0&0&0&0 \\
    0&0&0&0 \\
    0&0&0&0 \\
    0&0&0&0
  \end{smallmatrix}\right) , \\
  \bar V^{\perp2} &= -\tfrac{1071}{16} V^{\perp2} \\
    &= \left(\begin{smallmatrix}
    -318150&170280&249300&-18468 \\
    90027&-161100&94950&59778 \\
    0&51408&-420903&28917 \\
    -207088&422480&-1012475&-50895
  \end{smallmatrix}\right) , \\
  \bar V^{\perp4} &= -\tfrac{51}{64} V^{\perp4} \\
    &= \left(\begin{smallmatrix}
    -800100&170280&249300&-18468 \\
    90027&-643050&94950&59778 \\
    0&51408&-902853&28917 \\
    -207088&422480&-1012475&-532845
  \end{smallmatrix}\right) , \\
  \bar V^{\perp6} &= \tfrac{187}{102400} V^{\perp6} \\
    &= \left(\begin{smallmatrix}
    578277&170280&249300&-18468 \\
    90027&735327&94950&59778 \\
    0&51408&475524&28917 \\
    -207088&422480&-1012475&845532
  \end{smallmatrix}\right) , \\
  \bar V^{\perp8} &= \tfrac{2431}{278691840} V^{\perp8} \\
    &= \left(\begin{smallmatrix}
    647892&170280&249300&-18468 \\
    90027&804942&94950&59778 \\
    0&51408&545139&28917 \\
    -207088&422480&-1012475&915147
  \end{smallmatrix}\right) ,
\end{aligned}\end{equation}
where the fractions are determined by the condition that the entries
in each $\bar V^J$ be coprime integers. That the pairing force $V_0$
kills the $v = 4$ space $\Phi_{40}^\perp$ is no surprise. The matrices
$\bar V^{\perp J}$ exhibit the remarkable similarity
\begin{equation}\label{eq:displ}
   \bar V^{\perp2} = \bar V^{\perp4} + 481950
   = \bar V^{\perp6} - 896427 = \bar V^{\perp8} - 966042
\end{equation}
with multiplication of the scalars by the unit matrix understood. Like
the invariance of the entire $\Phi_{40}^\perp$, this regularity lacks
fundamental explanation. It follows that the interactions $\bar V_J$
represented on $\Phi_{40}^\perp$ by these matrices also act
identically on $\Phi_{40}^\perp$ except for these scalar terms. Any
linear combination of these interactions, that is, an arbitrary $V$,
then acts on $\Phi_{40}^\perp$ as a linear combination of any one of
them and a scalar. This applies, in particular, to the two-body
interaction $\I^2 - j(j+1)N$, where $N = \sum_m n_m$. Because $N$ acts
on the four-body space as the scalar 4, conversely then every $V_J$
acts on $\Phi_{40}^\perp$ as a linear combination of $\I^2$ and a
scalar. This explains, in particular, the zeros in the third row and
first column of every $V^{\perp J}$. For no $| m_1 m_2 m_3 m_4
\rangle$ in the expansion~\eqref{eq:chi} of $ \chi_1$, the expansion
of $\I^2 | m_1 m_2 m_3 m_4 \rangle$ on states $| m_1 m_2 m_3 m_4
\rangle$ indeed contains $\ba{9}{-1}{-3}{-5}$. Therefore $\chi_3$
cannot appear in the expansion of $\I^2 \chi_1$ on states $\chi_i$.

One arrives at a prediction that might be tested experimentally. To
the extend of validity of the $j = 9/2$ shell model, the spacings of
the energy levels with angular momenta $I$ and $I'$ must have the
ratio of $I(I+1)-I'(I'+1)$. The nucleus $^{94}$Ru has a closed neutron
major shell and 4 protons in the $1g_{9/2}$ subshell. The yrast $I=4$,
6, 10, and 12 levels (with tentative assignments $I=10$ and 12) have
excitation energies 2186.6, 2498.0, 3991.2, and
4716.6~keV~\cite{ref:NuDat}. The states with $I = 10$ and 12 are
expected to have fairly pure $1g_{9/2}$ configurations while,
according to Das~\textit{et~al.}~\cite{ref:Das22}, both multiplets
with $I = 4$ and 6 could be mixtures of those labeled $\gamma$ and
$\alpha$ in Sec.~\ref{sec:anal} due to perturbation by configurations
outside the proton $1g_{9/2}$ shell. The pure Escuderos-Zamick energy
levels should then be close to the observed yrast levels.
Extra\-polation by the spacing rule from $I = 10$ and 12 gives
\linebreak excitation energies 2571.9 and 2918.9~keV, somewhat above
the yrast levels. A similar analysis for $^{96}$Pd, with 4 holes in
the $1g_{9/2}$ shell (and tentative assignments of the angular momenta
concerned), predicts Escuderos-Zamick levels at 2237.9 and 2616.8~keV,
closer to the yrast levels at 2099.01 and 2424.19~keV. Interpreting
the second observed $I = 4$ and 6 levels in
$^{72}$Ni~\cite{ref:Mor16}, with a closed proton major shell and 4
neutrons in the $1g_{9/2}$ subshell, as Escuderos-Zamick levels leads
to the prediction of the $I = 10$ and 12 levels at 3603 and 4258~keV,
and the corresponding analysis using data from~\cite{ref:Mor18} gives
3750 and 4590~keV in $^{74}$Ni with 4 holes in the $1g_{9/2}$
subshell. The nucleus $^{212}$Rn has a closed neutron shell and 4
protons in the $1h_{9/2}$ subshell. Extrapolation from the observed
$I = 10$ and 12 levels with partly tentative spin
assignments~\cite{ref:NuDat} places the Escuderos-Zamick levels at
2212 and 2320~keV. This is close to observed levels at 2171.87 and
2324.43~keV with assignments $I = 6$ (tentative) and 4, but with these
assignments the order is the inverse of the one predicted. This may be
due to perturbation by levels outside the proton $1h_{9/2}$ subshell.
In $^{214}$Ra with 4 holes in the proton $1h_{9/2}$ subshell
extrapolation from the observed $I = 10$ and 12
levels~\cite{ref:NuDat} places the Escuderos-Zamick levels at 2333 and
2482~keV.

Since the operator $\I^2$ acts on the $M_I = 0$ space as $I_-I_+$, the
matrix $C$ representing its action on $\Phi_{40}^\perp$ is easily
calculated by \eqref{eq:chi}, \eqref{eq:ba}, and \eqref{eq:com}. By
comparison with~\eqref{eq:VJperpmat}, one finds in the notation of
\eqref{eq:displ} that
\begin{multline}
   \tfrac1{3213} \bar V^{\perp2} + 156
   = \tfrac1{3213} \bar V^{\perp4} + 306 \\
   = \tfrac1{3213} \bar V^{\perp6} - 123
   = \tfrac1{3213} \bar V^{\perp8} - \tfrac{434}3 = C .
\end{multline}
No simple expression in terms of $J$ seems to reproduce these
displacements. For $I = 4$ and 6, Van Isacker and Heinze calculated
the ratios $r^J_I = \mu^J_I/\nu_J$, where $\mu^J_I$ and $\nu_J$ are
the eigenvalues of $V_J$ in the four-fermion system and a two-fermion
state with $I = J$~\cite{ref:Isa08,ref:Isa14}.
From~\eqref{eq:V}--\eqref{eq:cm1m2}, one gets
\begin{equation}
  \nu_J = c_J^2 = \binom{2J}{J}^{-1} \sum_{m=-j}^j (c^J_{m,-m})^2 .
\end{equation}
My calculations confirm the values
$r^J_4=\frac{68}{33},1,\frac{13}{15},\frac{114}{55}$ and
$r^J_6=\frac{19}{11},\frac{12}{13},1,\frac{336}{143}$ for $J=2$--8
reported in~\cite{ref:Isa08,ref:Isa14}, and further provide $r^J_{10}
=\frac{23}{33},\frac{98}{143},\frac{233}{165},\frac{2292}{715}$
and $r^J_{12}=0,\frac{75}{143},\frac{93}{55},\frac{246}{65}$.

\section{\label{sec:actPhi}Action of  $V$ on $\Phi_{40}$}

Like in~\eqref{eq:VJperp}, the action of $V_J$ on the states $\psi_i$
may be expressed by matrices $V^J$. They are
\begin{widetext}\begin{equation}\begin{gathered}
  V^0 = 2 \left(\begin{smallmatrix}
    4&1&1&1&1&-2&-5&-5&1&7&7&1&-5&-5 \\
    1&4&1&1&1&-5&-2&9&3&-9&-9&3&9&-2 \\
    1&1&4&1&1&-5&9&-2&-6&0&0&-6&-2&9 \\
    1&1&1&4&1&1&3&-6&-2&0&0&-2&-6&3 \\
    1&1&1&1&4&7&-9&0&0&-2&-2&0&0&-9 \\
    0&0&0&0&0&0&0&0&0&0&0&0&0&0 \\
    0&0&0&0&0&0&0&0&0&0&0&0&0&0 \\
    0&0&0&0&0&0&0&0&0&0&0&0&0&0 \\
    0&0&0&0&0&0&0&0&0&0&0&0&0&0 \\
    0&0&0&0&0&0&0&0&0&0&0&0&0&0 \\
    0&0&0&0&0&0&0&0&0&0&0&0&0&0 \\
    0&0&0&0&0&0&0&0&0&0&0&0&0&0 \\
    0&0&0&0&0&0&0&0&0&0&0&0&0&0 \\
    0&0&0&0&0&0&0&0&0&0&0&0&0&0&
  \end{smallmatrix}\right) , \\
   V^2 = \tfrac{16}{3} \left(\begin{smallmatrix}
    630&108&-54&-162&-216&594&189&189&351&513&513&351&189&189 \\
    108&342&-18&-54&-72&189&594&567&405&81&81&405&567&594 \\
    -54&-18&315&27&36&189&567&594&162&324&324&162&594&567 \\
    -162&-54&27&387&108&351&405&162&594&324&324&594&162&405 \\
    -216&-72&36&108&450&513&81&324&324&594&594&324&324&81 \\
    18&0&0&0&0&522&45&-117&-171&-171&-171&-171&-117&45 \\
    7&9&0&0&0&79&412&-110&-152&-152&-152&-26&64&-143 \\
    5&0&9&0&0&32&-13&293&-133&-133&56&-7&-151&83 \\
    3&0&0&9&0&-15&-42&-96&282&111&-78&-60&12&12 \\
    1&0&0&0&9&-62&-71&-89&130&301&22&-59&94&-95 \\
    -1&0&0&0&9&-109&-100&107&-40&41&320&149&-76&-76 \\
    -3&0&0&9&0&-156&-3&51&-3&-21&168&339&-57&-57 \\
    -5&0&9&0&0&-149&58&-86&88&151&-38&-38&358&-38 \\
    -7&9&0&0&0&-34&-178&155&107&-19&-19&-19&-19&377
  \end{smallmatrix}\right) , \\
  V^4 = \tfrac{1344}{5} \left(\begin{smallmatrix}
    4518&-1188&-918&162&972&-20574&-6075&-6075&-12393&-18711&-18711&-12393&-6075&-6075 \\
    -1188&4998&1122&-198&-1188&-6075&-20574&-20817&-14499&-1863&-1863&-14499&-20817&-20574 \\
    -918&1122&4413&-153&-918&-6075&-20817&-20574&-5022&-11340&-11340&-5022&-20574&-20817 \\
    162&-198&-153&3573&162&-12393&-14499&-5022&-20574&-11340&-11340&-20574&-5022&-14499 \\
    972&-1188&-918&162&4518&-18711&-1863&-11340&-11340&-20574&-20574&-11340&-11340&-1863 \\
    -630&0&0&0&0&8730&1125&1395&585&-495&-495&585&1395&1125 \\
    -245&-315&0&0&0&655&8500&970&340&1000&1000&-290&-860&1585 \\
    -175&0&-315&0&0&320&395&8285&2345&245&-700&455&1115&-745 \\
    -105&0&0&-315&0&-15&30&2370&7590&615&120&390&570&-60 \\
    -35&0&0&0&-315&-350&1165&445&730&7225&1450&235&-470&1345 \\
    35&0&0&0&-315&-145&1220&-625&170&1505&7280&665&290&1040 \\
    105&0&0&-315&0&600&-435&105&195&285&780&7395&1905&-345 \\
    175&0&-315&0&0&1075&-1370&340&130&-425&520&2020&7510&-230 \\
    245&-315&0&0&0&470&710&-1945&-745&1385&1385&-115&-115&7625
  \end{smallmatrix}\right) , \\
   V^6 = \tfrac{1075200}{11} \left(\begin{smallmatrix}
    4221&-297&270&162&-216&21006&12339&12339&9585&6831&6831&9585&12339&12339 \\
    -297&5229&-990&-594&792&12339&21006&5913&8667&14175&14175&8667&5913&21006 \\
    270&-990&5040&540&-720&12339&5913&21006&12798&10044&10044&12798&21006&5913 \\
    162&-594&540&4464&-432&9585&8667&12798&21006&10044&10044&21006&12798&8667 \\
    -216&792&-720&-432&4716&6831&14175&10044&10044&21006&21006&10044&10044&14175 \\
    558&0&0&0&0&6057&-954&-387&1179&2475&2475&1179&-387&-954 \\
    217&279&0&0&0&-764&5959&1630&1264&472&472&634&328&364 \\
    155&0&279&0&0&-736&1253&6338&728&1169&224&-469&323&476 \\
    93&0&0&279&0&183&426&291&6501&417&714&192&-816&237 \\
    31&0&0&0&279&1102&-617&445&70&6334&493&367&-470&-470 \\
    -31&0&0&0&279&1373&-364&-427&236&218&6059&-61&488&-511 \\
    -93&0&0&279&0&996&555&-687&-201&-111&-408&6108&420&744 \\
    -155&0&279&0&0&349&1006&538&-1124&-1151&-206&73&6553&1783 \\
    -217&279&0&0&0&-190&1106&629&-283&-1453&-1453&347&1931&6701
  \end{smallmatrix}\right) , \\
  V^8 = \tfrac{975421440}{143} \left(\begin{smallmatrix}
    7143&-21&60&-84&42&-12342&-8823&-8823&-4845&-867&-867&-4845&-8823&-8823 \\
    -21&7287&-420&588&-294&-8823&-12342&459&-3519&-11475&-11475&-3519&459&-12342 \\
    60&-420&8340&-1680&840&-8823&459&-12342&-9486&-5508&-5508&-9486&-12342&459 \\
    -84&588&-1680&9492&-1176&-4845&-3519&-9486&-12342&-5508&-5508&-12342&-9486&-3519 \\
    42&-294&840&-1176&7728&-867&-11475&-5508&-5508&-12342&-12342&-5508&-5508&-11475 \\
    -306&0&0&0&0&4491&-72&9&-1053&-1845&-1845&-1053&9&-72 \\
    -119&-153&0&0&0&-122&5657&-2530&-1068&-584&-584&-438&-76&-1078 \\
    -85&0&-153&0&0&512&-2111&6184&-2436&-973&-28&273&-491&-542 \\
    -51&0&0&-153&0&-141&-342&-2097&6723&-2019&-588&-234&522&-369 \\
    -17&0&0&0&-153&-794&239&-445&-1770&7192&-2321&-339&470&20 \\
    17&0&0&0&-153&-1051&28&469&-222&-2116&7397&-1653&-446&247 \\
    51&0&0&-153&0&-912&-345&519&117&27&-1404&7074&-2100&-318 \\
    85&0&-153&0&0&-503&-502&-496&858&997&52&-1851&6179&-2071 \\
    119&-153&0&0&0&50&-1022&-83&381&851&851&-249&-2537&5713
  \end{smallmatrix}\right) .
\end{gathered}\end{equation}\end{widetext}
Some patterns leap to the eye. The entries in $V^0$ are easily
understood. Thus $V_0 |m_1 m_2 m_3 m_4\rangle$ vanishes unless two of
the $m$'s form a pair $m,-m$, in which case the remaining two do the
same. Further,
$V_0 |m_1,m_2,-m_2,-m_1\rangle = (-)^{m_1-m_2+1} 2 (n_{m_1} + n_{m_2})
\phi_0$, so the states $n_m \phi_1$ do not contribute to any
$V_0 \psi_i$. This expression follows from
$V_0 |m,-m\rangle = 2 \sum_{m'} (-)^{m'-m}|m',-m'\rangle$, where
$|m_1 m_2\rangle = \alpha_{m_1}^\dagger \alpha_{m_2}^\dagger
|\rangle$, and the observation that the coefficient of
$|m_1,m_2,-m_2,-m_1\rangle$ in the expansion~\eqref{eq:phi} of
$\phi_0$ is $(-1)^{m_1-m_2+1}$. By using it in combination with the
expansions~\eqref{eq:phi}, it is, in fact, straightforward to
reconstruct every entry in $V^0$, and in particular, the simple
pattern in its upper left $5 \times 5$ submatrix. Notice to this end
that the last eight terms in the expansion of $\phi_1$ do not
contribute to $V_0 \phi_1$.

For a general $J$, one notices in the upper right $5 \times 8$
submatrix of $V^J$ equal contributions to $V_J n_{\pm m} \phi_1$ from
any $n_{m'} \phi_0$. This is an immediate consequence of
$n_{m'} \phi_0 = n_{-m'} \phi_0$ and the symmetry under half-turn
rotations about an axis perpendicular to the quantization axis. The
same pattern is seen in the parts of the sixth rows just below, which
display contributions to $V_J n_m \phi_1$ from $n_j \phi_1$ for
$m \ne \pm j$, and again the reason is the symmetry under half-turn
rotations about an axis perpendicular to the quantization axis. Such a
rotation thus leads to both the replacement of $m$ by $-m$ and the
omission of $n_j \phi_1$ instead of $n_{-j} \phi_1$ in the selection
of the states $\psi_i$. But by $\sum_m m n_m \phi_1 = J_0 \phi_1 = 0$,
the state $n_{-j} \phi_1$ equals $n_j \phi_1$ plus a linear
combination of states that are common to both the original and the new
basis. Therefore in the original basis, the contribution of
$n_j \phi_1$ to $V_J n_{-m} \phi_1$ equals its contribution to
$V_J n_m \phi_1$. It follows further that when the lower right
$9 \times 9$ submatrix of $V^J$ is written $(v^{1J}_{mm'}|m,m' > -j)$
with indices referring to the basic states $n_m \phi_1$, then for
$m,m' < j$, one should have
\begin{equation}
  v^{1J}_{-m',-m} = v^{1J}_{m'm} - \frac{m'}j v^{1J}_{jm} .
\end{equation}
This is verified by inspection. Similar patterns occur when a state
$n_m \phi_1$ other than $n_{-j} \phi_1$ is omitted in the selection of
the states $\psi_i$.

It is trivial by $n_m \phi_0 = n_{-m} \phi_0$ that the state
$n_{m'} \phi_0$ contributes equally to $V_J n_{\pm m} \phi_0$.
Therefore when $v^{00J}_{mm'}$ denotes the entries in the
upper left $5 \times 5$ submatrix of $V^J$ with indices referring to
the basic states $n_m \phi_0$, and $v^{01J}_{mm'}$ those of its
neighboring $5 \times 5$ submatrix to the right with indices referring
to the basic states $n_m \phi_0$ and $n_{m'} \phi_1$, then for both
$i = 0$ and 1 one has
\begin{multline}\label{eq:V0iJm}
  \tfrac12 V_J \bar n_m \phi_i
    = \sum_{m'>0} v^{0iJ}_{m'm} \bar n_{m'} \phi_0 \\
    + \text{linear combination of $n_{m'} \phi_1$}
\end{multline}
with $\bar n_m = n_m + n_{-m}$. The operators $\bar n_m$ can be
expressed by the tensor operators $T_{I0}$ defined in the
introduction,
\begin{equation}
  \bar n_m = (-)^{j-m} 2 \sum_\text{even $I$}
    \langle jmj-m|I0\rangle T_{I0} .
\end{equation}
Since the states $T_{I0} \phi_i$ span the subspace of $\Phi_{40}$ with
angular momentum $I$, which is invariant to $V^J$, one can write
\begin{equation}\label{eq:V0iJI}
  V_J T_{I0} \phi_i = \sum_{i'} w^{IJ}_{i'i} T_{I0} \phi_{i'} .
\end{equation}
By combining the equations~\eqref{eq:V0iJm}--\eqref{eq:V0iJI} and the
definition of $T_{I0}$ in the introduction one obtains
\begin{multline}
  v^{0iJ}_{mm'} = (-)^{m-m'} \sum_\text{even $I$} \\
    \langle jmj-m|I0\rangle
      \langle jm'j-m'|I0\rangle w^{IJ}_{0i} ,
\end{multline}
which is symmetric in $m$ and $m'$. This explains that the said
submatrices of every $V^J$ are symmetric.

A spectacular pattern emerges in the lower left $9 \times 5$ submatrix
of every $V^J$. For $j > m \ge 1/2$ the only contribution to
$V_J n_m \phi_0$ from states $n_{m'} \phi_1$ is a term
$\gamma_J \bar n_m \phi_1$, where $\gamma_J$ is constant. Upon closer
inspection taking into account again $\sum_m m n_m \phi_1 = 0$, this
also hold for $m = j$. For every $m$ one thus has
\begin{equation}\label{eq:Vjphi0m}
  \tfrac12 V_J \bar n_m \phi_0
    = \sum_{m'>0} v^{00J}_{m'm} \bar n_{m'}  \phi_0
    + \gamma_J \bar n_m \phi_1 .
\end{equation}
(The states $\bar n_m \phi_0$ span the $v \le 2$
subspace of $\Phi_{40}$.) This is not dependent on the choice of
$\phi_1$. Replacing $\phi_1$ by any linear combination
$\phi_1 + \epsilon \phi_0$ gives a relation of the same structure with
the same $\gamma_J$. In particular, $\phi_1$ could have $v = 4$. The
origin of this pattern is explained in Appendix~\ref{sec:(25)}.

\section{\label{sec:con}Conclusion}

The existence of the Escuderos-Zamick states was shown to be
equivalent to the invariance to any rotationally invariant two-body
interaction of the span $\Phi_4$ of states generated from angular
momentum zero states by one-body operators. This equivalence explains
the occurrence of the Escuderos-Zamick states for exactly the angular
momenta 4 and~6. The said property of the angular momentum zero state
space was verified by exact calculation. This verification was
facilitated by the observation that it is required only for the
subspace $\Phi_{40}$ of $\Phi_4$ characterized by magnetic quantum
number $M_I = 0$ or its orthogonal complement $\Phi_{40}^\perp$ within
the $M_I = 0$ space. The actions of five basic rotationally invariant
two-body interactions on 4 basic states in $\Phi_{40}^\perp$ and 14
basic states in $\Phi_{40}$ were displayed in matrix form, and
remarkable regularities in these matrices disclosed. One of them leads
to a rule that relates the Escuderos-Zamick energy levels to levels
with $I = 10$ and 12. This rule was applied in a discussion of certain
aspects of the spectra of $^{94}$Ru, $^{96}$Pd, and $^{74}$Ni. The
said regularity so far lacks fundamental explanation. Understanding it
could possibly provide a clue towards a more intuitive understanding
of the invariance of $\Phi_4$.

\appendix

\section{\label{sec:cmm}Proof that $c^J_{m_1m_2}$ are integral}

First notice that the algorithm for $c^J_{m_1m_2}$ described in
Sec.~\ref{seq:V} ensures the proportionality~\eqref{eq:cm1m2} so that
$P_{JM}^\dagger$ given by~\eqref{eq:P+a} is a tensor operator.
Besides~\eqref{eq:rec-},
$[I_+,P_{JM}^\dagger] = \sqrt{(J-M)(J+M+1)} P_{J,M+1}^\dagger$,
\eqref{eq:P+al}, and \eqref{eq:com} give
\begin{equation}\label{eq:rec+}
  (J + m_1 + m_2) c^J_{m_1m_2} = c^J_{m_1-1,m_2} + c^J_{m_1,m_2-1} .
\end{equation}
With
\begin{equation}\label{eq:d}
  d^J_{m_1m_2} = (J - m_1 -m_2)! \, c^J_{m_1m_2} ,
\end{equation}
the recursion relations \eqref{eq:rec-} and \eqref{eq:rec+} take the
forms
\begin{gather}\label{eq:drec-}\begin{split}
    d^J_{m_1m_2} = (j - m_1)(j + m_1 + 1) d^J_{m_1+1,m_2} \\
      + (j - m_2)(j + m_2 + 1) d^J_{m_1,m_2+1} ,
\end{split} \\
\label{eq:drec+}\begin{split}
    (J + m_1 + m_2)(J - m_1 - m_2 + 1) d^J_{m_1m_2} \\
    = d^J_{m_1-1,m_2} + d^J_{m_1,m_2-1} .
\end{split}\end{gather}
(Again, terms with undefined values of $d^J_{m_1'm_2'}$ are omitted.)
Setting $m_1 = j$ so that only the second term occurs on the right
in~\eqref{eq:drec-}, and using also $d^J_{j,J-j} = 1$, one gets for
$J - j \ge m_2 \ge -j$, by repeated application of~\eqref{eq:drec-},
an expression for $d^J_{jm_2}$ as a product of two products of
$J - j - m_2$ consecutive integers. Hence $d^J_{jm_2}$ is divisible by
$(J - j - m_2)!^2$, and, all the more, by $(J - j - m_2)!$. It then
follows by induction by means of~\eqref{eq:drec+} that
$(J - m_1 - m_2)!$ divides $d^J_{m_1m_2}$ for every $m_1,m_2$ with
$m_1 + m_2 \ge 0$. Then by~\eqref{eq:d}, $c^J_{m_1m_2}$ is integral
for $m_1 + m_2 \ge 0$. For $m_1 + m_2 < 0$ one can now apply the
second equation in~\eqref{eq:csym}. In this case, $j + m_1 < j - m_2$
and $j + m_2 < j - m_1$, so the first factor on the right, and hence
$c^J_{m_1m_2}$, are integral.

\section{\label{sec:(25)}Explanation of~\eqref{eq:Vjphi0m}}

The first equation in~\eqref{eq:phi} can be written
\begin{equation}
  \phi_0 = - \tfrac12 {P^\dagger}^2 | \rangle
\end{equation}
with
\begin{equation}\label{eq:P+}
  P^\dagger = \sum_{m>0} (-)^{j-m} a_m^\dagger a_{-m}^\dagger .
\end{equation}
Hence
\begin{multline}
  n_m \phi_0 = a_m^\dagger a_m \phi_0  \\
  = - a_m^\dagger P^\dagger [ a_m , P^\dagger ] | \rangle
  = - (-)^{j-m} P^\dagger a_m^\dagger a_{-m}^\dagger | \rangle ,
\end{multline}
and
\begin{multline}\label{eq:Vnphi}
  V_J n_m \phi_0
  = \tfrac12 \sum_M P_{JM}^\dagger P_{JM} n_m \phi_0 \\
  = - (-)^{j-m} \tfrac12 \sum_M P_{JM}^\dagger
    ( P^\dagger P_{JM} + [ P_{JM} , P^\dagger ] )
    a_m^\dagger a_{-m}^\dagger | \rangle .
\end{multline}
Since $P_{JM} a_m^\dagger a_{-m}^\dagger | \rangle \propto | \rangle$,
the first term in the parentheses contributes to $V_J n_m \phi_0$ a
term proportional to $P_{JM}^\dagger P^\dagger | \rangle$, which has
$v \le 2$. I proceed by calculating the commutator
$[ P_{JM} , P^\dagger]$.

For convenience, I omit for now the factor $c_J$ in~\eqref{eq:P+a}. It
is reentered at the end of this appendix. I then have

\newpage

\noindent\begin{minipage}{\columnwidth}

\begin{equation}\label{eq:[P,P]}\begin{aligned}[]
  [ P_{JM} , & \, P^\dagger ] = \tfrac12 \sum_{m_1m_2m} (-)^{j-m} \\
    & \qquad \langle j m_1 j m_2 | J M \rangle
    [ a_{m_2} a _{m_1} , a_m^\dagger a_{-m}^\dagger ] \\
  = & \sum_{m_1m_2} \langle j m_1 j m_2 | J M \rangle \\
    & \qquad ( (-)^{j-m_1} a_{m_2} a _{-m_1}^\dagger
      + (-)^{j-m_2} a_{-m_2}^\dagger a_{m_1} ) \\
  = & \sum_{m_1m_2} (-)^{j-m_2} \langle j m_1 j m_2 | J M \rangle
      [ a_{-m_2}^\dagger , a_{m_1} ] \\
  = & \ 2 \sum_{m_1m_2} (-)^{j-m_2} \langle j m_1 j m_2 | J M \rangle
      a_{-m_2}^\dagger a_{m_1} \\
    & - \sum_m (-)^{j+m} \langle j m j \!-\! m | J M \rangle \\
  = & \ 2 \sum_{m_1m_2} (-)^{j-m_2} \langle j m_1 j m_2 | J M \rangle
      a_{-m_2}^\dagger a_{m_1} \\
    & + \delta_{J0} \sqrt{2 j + 1} .
\end{aligned}\end{equation}
Since $P_{J0}^\dagger \propto P^\dagger$, the last term in this
expression is seen by comparison with~\eqref{eq:P+} to contribute
in~\eqref{eq:Vnphi} a term proportional to $n_m \phi_0$. 

\quad Further,
\begin{equation}
  a_{-m_2}^\dagger a_{m_1} a_m^\dagger a_{-m}^\dagger | \rangle
  = ( \delta_{m_1m} - \delta_{m_1-m} )
    a_{-m_2}^\dagger a_{-m_1}^\dagger  | \rangle ,
\end{equation}
so the first term in the expression~\eqref{eq:[P,P]} contributes to
the sum in~\eqref{eq:Vnphi} terms
\begin{multline}
  2 a_m^\dagger \sum_{Mm_2} (-)^{j-m_2}
    \langle j \!-\! m j m_2 | J M \rangle P_{JM}^\dagger
     a_{-m_2}^\dagger | \rangle \\
    - \text{same with $-m$ instead of $m$} .
\end{multline}
Here~\cite{ref:Edm57},
\begin{multline}
  \xi = \sum_{Mm_2} (-)^{j-m_2}
    \langle j \!-\! m j m_2 | J M \rangle
    P_{JM}^\dagger a_{-m_2}^\dagger  | \rangle \\
  = - \sqrt{\frac{2 J+1}{2 j+1}}
    \sum_{m_2M} \langle j \!-\! m_2 J M | j \!-\! m \rangle
    a_{-m_2}^\dagger P_{JM}^\dagger | \rangle
\end{multline}
is a member with $M_I = -m$ of the space $\Phi_3$ defined in
Sec.~\ref{sec:anal}. With standard relative phases within its
multiplet, one therefore has
\begin{equation}
  \xi = (-)^{j+m} a_m (\zeta \phi_0 + \eta \phi_1 ) ,
\end{equation}
where $\zeta$ and $\eta$ do not depend on $m$. Totally, one arrives at
\begin{equation}
  V_J n_m \phi_0  = \eta (n_m + n_{-m})  \phi_1
    + \text{$v \le 2$ state} ,
\end{equation}
which is equivalent to~\eqref{eq:Vjphi0m} with $\gamma_J = \eta c_J^2$.

\end{minipage}

\newpage

\bibliography{escuderos}

\begin{thebibliography}{11}%
\makeatletter
\providecommand \@ifxundefined [1]{%
 \@ifx{#1\undefined}
}%
\providecommand \@ifnum [1]{%
 \ifnum #1\expandafter \@firstoftwo
 \else \expandafter \@secondoftwo
 \fi
}%
\providecommand \@ifx [1]{%
 \ifx #1\expandafter \@firstoftwo
 \else \expandafter \@secondoftwo
 \fi
}%
\providecommand \natexlab [1]{#1}%
\providecommand \enquote  [1]{``#1''}%
\providecommand \bibnamefont  [1]{#1}%
\providecommand \bibfnamefont [1]{#1}%
\providecommand \citenamefont [1]{#1}%
\providecommand \href@noop [0]{\@secondoftwo}%
\providecommand \href [0]{\begingroup \@sanitize@url \@href}%
\providecommand \@href[1]{\@@startlink{#1}\@@href}%
\providecommand \@@href[1]{\endgroup#1\@@endlink}%
\providecommand \@sanitize@url [0]{\catcode `\\12\catcode `\$12\catcode
  `\&12\catcode `\#12\catcode `\^12\catcode `\_12\catcode `\%12\relax}%
\providecommand \@@startlink[1]{}%
\providecommand \@@endlink[0]{}%
\providecommand \url  [0]{\begingroup\@sanitize@url \@url }%
\providecommand \@url [1]{\endgroup\@href {#1}{\urlprefix }}%
\providecommand \urlprefix  [0]{URL }%
\providecommand \Eprint [0]{\href }%
\providecommand \doibase [0]{http://dx.doi.org/}%
\providecommand \selectlanguage [0]{\@gobble}%
\providecommand \bibinfo  [0]{\@secondoftwo}%
\providecommand \bibfield  [0]{\@secondoftwo}%
\providecommand \translation [1]{[#1]}%
\providecommand \BibitemOpen [0]{}%
\providecommand \bibitemStop [0]{}%
\providecommand \bibitemNoStop [0]{.\EOS\space}%
\providecommand \EOS [0]{\spacefactor3000\relax}%
\providecommand \BibitemShut  [1]{\csname bibitem#1\endcsname}%
\let\auto@bib@innerbib\@empty
\bibitem [{\citenamefont {Escuderos}\ and\ \citenamefont
  {Zamick}(2006)}]{ref:Esc06}%
  \BibitemOpen
  \bibfield  {author} {\bibinfo {author} {\bibfnamefont {A.}~\bibnamefont
  {Escuderos}}\ and\ \bibinfo {author} {\bibfnamefont {L.}~\bibnamefont
  {Zamick}},\ }\href@noop {} {\bibfield  {journal} {\bibinfo  {journal} {Phys.
  Rev. C}\ }\textbf {\bibinfo {volume} {73}},\ \bibinfo {pages} {044302}
  (\bibinfo {year} {2006})}\BibitemShut {NoStop}%
\bibitem [{\citenamefont {Talmi}(2010)}]{ref:Tal10}%
  \BibitemOpen
  \bibfield  {author} {\bibinfo {author} {\bibfnamefont {I.}~\bibnamefont
  {Talmi}},\ }\href@noop {} {\bibfield  {journal} {\bibinfo  {journal} {Nucl.
  Phys. A}\ }\textbf {\bibinfo {volume} {846}},\ \bibinfo {pages} {31}
  (\bibinfo {year} {2010})}\BibitemShut {NoStop}%
\bibitem [{\citenamefont {{Van Isacker}}\ and\ \citenamefont
  {Heinze}(2008)}]{ref:Isa08}%
  \BibitemOpen
  \bibfield  {author} {\bibinfo {author} {\bibfnamefont {P.}~\bibnamefont {{Van
  Isacker}}}\ and\ \bibinfo {author} {\bibfnamefont {S.}~\bibnamefont
  {Heinze}},\ }\href@noop {} {\bibfield  {journal} {\bibinfo  {journal} {Phys.
  Rev. Lett.}\ }\textbf {\bibinfo {volume} {100}},\ \bibinfo {pages} {052501}
  (\bibinfo {year} {2008})}\BibitemShut {NoStop}%
\bibitem [{\citenamefont {{Van Isacker}}\ and\ \citenamefont
  {Heinze}(2014)}]{ref:Isa14}%
  \BibitemOpen
  \bibfield  {author} {\bibinfo {author} {\bibfnamefont {P.}~\bibnamefont {{Van
  Isacker}}}\ and\ \bibinfo {author} {\bibfnamefont {S.}~\bibnamefont
  {Heinze}},\ }\href@noop {} {\bibfield  {journal} {\bibinfo  {journal} {Ann.
  Phys. (N. Y.)}\ }\textbf {\bibinfo {volume} {349}},\ \bibinfo {pages} {73}
  (\bibinfo {year} {2014})}\BibitemShut {NoStop}%
\bibitem [{\citenamefont {Qi}(2017)}]{ref:Qi17}%
  \BibitemOpen
  \bibfield  {author} {\bibinfo {author} {\bibfnamefont {C.}~\bibnamefont
  {Qi}},\ }\href@noop {} {\bibfield  {journal} {\bibinfo  {journal} {Phys.
  Lett. B}\ }\textbf {\bibinfo {volume} {773}},\ \bibinfo {pages} {616}
  (\bibinfo {year} {2017})}\BibitemShut {NoStop}%
\bibitem [{\citenamefont {Qi}\ \emph {et~al.}(2012)\citenamefont {Qi},
  \citenamefont {Xu},\ and\ \citenamefont {Liotta}}]{ref:Qi12}%
  \BibitemOpen
  \bibfield  {author} {\bibinfo {author} {\bibfnamefont {C.}~\bibnamefont
  {Qi}}, \bibinfo {author} {\bibfnamefont {Z.~X.}\ \bibnamefont {Xu}}, \ and\
  \bibinfo {author} {\bibfnamefont {R.~J.}\ \bibnamefont {Liotta}},\
  }\href@noop {} {\bibfield  {journal} {\bibinfo  {journal} {Nucl. Phys. A}\
  }\textbf {\bibinfo {volume} {884-885}},\ \bibinfo {pages} {21} (\bibinfo
  {year} {2012})}\BibitemShut {NoStop}%
\bibitem [{\citenamefont {Edmonds}(1957)}]{ref:Edm57}%
  \BibitemOpen
  \bibfield  {author} {\bibinfo {author} {\bibfnamefont {A.~R.}\ \bibnamefont
  {Edmonds}},\ }\href@noop {} {\emph {\bibinfo {title} {Angular Momentum in
  Quantum Mechanics}}}\ (\bibinfo  {publisher} {Princeton University Press,
  Princeton, USA},\ \bibinfo {year} {1957})\BibitemShut {NoStop}%
\bibitem [{\citenamefont {{National Nuclear Data Center, NuDat
  3.0}}(2022)}]{ref:NuDat}%
  \BibitemOpen
  \bibfield  {author} {\bibinfo {author} {\bibnamefont {{National Nuclear Data
  Center, NuDat 3.0}}},\ }\href@noop {} {}\bibinfo {howpublished}
  {\texttt{https://www.nndc.bnl.gov/nudat3}} (\bibinfo {year} {retrieved June
  2022})\BibitemShut {NoStop}%
\bibitem [{\citenamefont {{Das \textit{et al.}}}(2022)}]{ref:Das22}%
  \BibitemOpen
  \bibfield  {author} {\bibinfo {author} {\bibfnamefont {K.}~\bibnamefont {{Das
  \textit{et al.}}}},\ }\href@noop {} {\bibfield  {journal} {\bibinfo
  {journal} {Phys. Rev. C}\ }\textbf {\bibinfo {volume} {105}},\ \bibinfo
  {pages} {L031304} (\bibinfo {year} {2022})}\BibitemShut {NoStop}%
\bibitem [{\citenamefont {{Morales \textit{et al.}}}(2016)}]{ref:Mor16}%
  \BibitemOpen
  \bibfield  {author} {\bibinfo {author} {\bibfnamefont {A.~I.}\ \bibnamefont
  {{Morales \textit{et al.}}}},\ }\href@noop {} {\bibfield  {journal} {\bibinfo
   {journal} {Phys. Rev. C}\ }\textbf {\bibinfo {volume} {93}},\ \bibinfo
  {pages} {034328} (\bibinfo {year} {2016})}\BibitemShut {NoStop}%
\bibitem [{\citenamefont {{Morales \textit{et al.}}}(2018)}]{ref:Mor18}%
  \BibitemOpen
  \bibfield  {author} {\bibinfo {author} {\bibfnamefont {A.~I.}\ \bibnamefont
  {{Morales \textit{et al.}}}},\ }\href@noop {} {\bibfield  {journal} {\bibinfo
   {journal} {Phys. Lett. B}\ }\textbf {\bibinfo {volume} {781}},\ \bibinfo
  {pages} {706} (\bibinfo {year} {2018})}\BibitemShut {NoStop}%
\end{thebibliography}%

\end{document}